\documentclass[12pt,a4paper]{article}
\usepackage{graphicx,color,amsmath, amssymb,cite,multirow}
\usepackage[english]{babel}
\usepackage{t1enc}
\usepackage[latin2]{inputenc}
\begin{document}
\textwidth=135mm
 \textheight=200mm

\begin{center}
{\bfseries Initial temperature and EoS of quark matter from direct photons
\footnote{{\small Talk at the VI Workshop on Particle Correlations and Femtoscopy, Kiev, September 14-18, 2010.}}}
\vskip 5mm
M. Csanád$^\dag$ I. Májer$^\dag$
\vskip 5mm
{\small {\it $^\dag$ Eötvös University, Department of Atomic Physics, Pázmány Péter s. 1/A, H-1117 Budapest, Hungary}}
\\
\end{center}
\vskip 5mm
\centerline{\bf Abstract}
The time evolution of the quark gluon plasma created in gold-gold collisions of the Relativistic Heavy Ion Collider (RHIC)
can be described by hydrodynamical models. Distribution of hadrons reflects the freeze-out state of
the matter. To investigate the time evolution one needs to analyze penetrating probes, such as direct
photon spectra. Distributions of low energy photons was published in 2010 by PHENIX. 
In this paper we analyze a 3+1 dimensional solution of relativistic hydrodynamics and calculate momentum distribution
of direct photons. Using earlier fits of this model to hadronic spectra, we compare photon calculations to
measurements and find that the initial temperature of the center of the fireball
is at least 519$\pm$12 MeV, while for the equation of state we get $c_s= 0.36\pm0.02$.
\vskip 10mm

\section{Perfect fluid hydrodynamics}
In the last several years it has been revealed that the matter produced in the collisions of the Relativistic Heavy Ion Collider (RHIC)
is a nearly perfect fluid~\cite{Lacey:2006bc}, i.e.\ it can be described with perfect fluid hydrodynamics. There was a long search for exact hydrodynamic models
(solutions of the partial differential equations of hydrodynamics) and several models proved to be applicable. There are 1+3 dimensional models,
 as well as relativistic models - but no 1+3 dimensional \emph{and} relativistic model has been tested yet. In this paper
we extract observables from the relativistic, ellipsoidally symmetric model of ref.~\cite{Csorgo:2003ry}. Hadronic observables 
were calculated in ref.~\cite{Csanad:2009wc}. Here we calculate momentum distribution and elliptic flow of direct photons.

Perfect fluid hydrodynamics is based on local conservation of
entropy or number density ($n$), energy-momentum density (chosen as $T^{\mu\nu}=(\epsilon+p)u^\mu u^\nu-pg^{\mu\nu}$) for perfect fluids.
Here $\epsilon$ is energy density, $p$ is pressure and $g^{\mu\nu}$ is the metric tensor, diag$(-1,1,1,1)$. The conservation equations
are closed by the equation of state, which gives the relationship between $\epsilon$ and $p$. Typically $\epsilon = \kappa p$ is chosen.
Note that exact, analytic result for hydrodynamic models is, that the hadronic observables do not depend on the initial state or the
equation of state separately, just through the final state~\cite{Csanad:2009sk}. Thus if we fix the final state from the data, and
determine initial state parameters from direct photon spectra.

Even though many solve the above equations numerically, there are only a few exact solutions for these equations.
See a small review of solutions in ref.~\cite{Csanad:2009wc}. As far as we know, until now there was only one 1+3
dimensional relativistic solutions investigated: the solution in ref.~\cite{Csorgo:2003ry}.
Hadronic observables from this solution were computed and compared to data in ref.~\cite{Csanad:2009wc}. Present
paper calculates thermal photon observables from this realistic 1+3 dimensional model and compares them to data for
the first time. Our method is different from numerical calculations: here one can determine the best values of the
parameters of the solution by fitting the analytic model results to data.

\section{The analyzed solution}
The analyzed solution, as described in refs.~\cite{Csorgo:2003ry,Csanad:2009wc} assumes self-similarity and ellipsoidal symmetry.
The ellipsoidal symmetry means that at a given proper time the thermodynamical quantities are constant on the surface of expanding ellipsoids.
The ellipsoids are given by constant values of the scale variable $s=r_x^2/X(t)^2+r_y^2/Y(t)^2+r_z^2/Z(t)^2$, where $X(t)=\dot X_0t$, $Y(t)=\dot Y_0t$,
and $Z(t)=\dot Z_0t$ are time proportional scale parameters (axes of the $s=1$ ellipsoid), and spatial coordinates are $r_x$, $r_y$, and $r_z$.
The velocity-field is a 3D Hubble-type expansion, $u^\mu(x)=x^\mu/\tau$. This means that the solution is accelerationless.
Note that similarly to ref.~\cite{Csanad:2009wc}, we use transverse expansion $u_t$ and eccentricity $\epsilon$ instead of $x$ and $y$ direction
expansion rates $\dot X_0$ and $\dot Y_0$.

The temperature is $T(x)=T_0\left(\tau_0/\tau\right)^{3/\kappa} \nu(s)^{-1}$, the number density $n(x)=n_0\left(\tau_0/\tau\right)^3 \nu(s)$
with $\tau$ being the proper time and $\nu(s)$ an arbitrary function of $s$. We choose $\tau_0$ to be the time of the freeze-out,
thus $T_0$ is the central freeze-out temperature. The function $\nu(s)$ is chosen as $\nu(s)=\exp(-bs/2)$, where $b = \left.\frac{\Delta T}{T}\right|_r$
is the temperature gradient. If the fireball is the hottest in the center, then $b<0$.
Note that for the momentum distribution of direct photons, we don't need any kind of density just the temperature distribution.

The picture used in hydro models is that the pre freeze-out (FO) medium is described by hydrodynamics, and the post freeze-out medium
is that of observed hadrons. The hadronic observables can be extracted from the solution via the phase-space distribution at the
FO. This will correspond to the hadronic final state or source distribution $S(x,p)$. See details about this topic in ref.~\cite{Csanad:2009wc}.
It is important to see that the same final state can be achieved with different equations of state or initial conditions~\cite{Csanad:2009sk}.
However, as we will see below, the source function of photons is sensitive to the whole time evolution, thus both to initial conditions
and equation of state as well.

For the source function of photon creation we have then
\begin{align}
S(x,p)d^4x = \mathcal{N}\frac{p_{\mu}u^{\mu}(x)}{\exp\left(p_{\mu}u^{\mu}(x)/T(x)\right)-1}\,d^4x
 \end{align}
where $\mathcal{N}=g/(2\pi)^3$ (with degeneracy factor $g$), $p_\mu u^\mu$ is the energy of the photon
in the co-moving system (from the Cooper-Frye prefactor). We will use a second order saddle-point approximation.
In this approximation the point of maximal emittivity and Gaussian source widths can be calculated.

\section{Thermal photon observables}
The invariant single particle momentum distribution can then be calculated from the source function by integrating on space-time.
Our calculated quantity will be $N_1(p_t)$, which is the invariant single particle momentum distribution taken at midrapidity
and integrated on the azimuthal angle $\varphi$, similarly to ref.~\cite{Csanad:2009wc}.
The result on the invariant transverse momentum distribution depends on the initial and final times. We introduce the variable $\xi=t/\tau_0$
(with $\tau_0$ as the freeze-out time), then in terms of $\xi$ the time integration goes from $i$ to $1$, if $i=t_i/\tau_0$. The result is: 
\begin{align}\label{e:N1ptfinal}
&N_1(p_t) = \sum_{n=0}^{\infty}(2\pi)^{\frac{3}{2}}\sqrt{\rho_x\rho_y\rho_z}\tau_0^4T_0
\left(\frac{p_t}{T_0}\right)^{\frac{3-4\kappa}{3}}\frac{\kappa}{3}\frac{B^n}{A^{n+\frac{4\kappa}{3}-\frac{3}{2}}}\nonumber\\
&\Bigg\lbrace  \left[ \frac{(\rho_x-1)^2+(\rho_y-1)^2}{4}(a_{0n}+a_{1n}) + \frac{a_{0n}-a_{1n}}{4} \right]\times  \nonumber\\
& \left.\Gamma\left(n + \frac{4\kappa}{3}-\frac{3}{2}, A\frac{p_t}{T_0}\xi^{\frac{3}{\kappa}} \right) \right|_1^i + \nonumber \\
& \left. \frac{\rho_x^2 + \rho_y^2 + \rho_z^2}{2} \,a_{0n} A \left.\Gamma\left(n + \frac{4\kappa}{3}-\frac{5}{2}, A\frac{p_t}{T_0}\xi^{\frac{3}{\kappa}} \right) \right|_1^i \right\rbrace\textnormal{.}
\end{align}
where we introduced the auxiliary quantities $\rho_x = \kappa/\left(\kappa -3-\kappa b/\dot{X_0^2}\right)$,
and $\rho_y$, $\rho_z$ similarly, while $A = 1 - (\rho_x + \rho_y)/4$ and
$B = (\rho_x - \rho_y)/4$. Furthermore, $a_0$, $a_1$ are the Taylor-coefficients of the first two modified Bessel functions:
\begin{align}
I_0(x) = \sum_{n=0}^{\infty} \! a_{0n}x^n&\quad\textnormal{   with   }\quad a_{0} = \left(1, 0, \frac{1}{4}, 0, \frac{1}{64}, 0, \frac{1}{2304}, 0, \, ...\right)\\
I_1(x) = \sum_{n=0}^{\infty} \! a_{1n}x^n&\quad\textnormal{   with   }\quad a_{1} = \left(0, \frac{1}{2}, 0, \frac{1}{16}, 0, \frac{1}{384}, 0, \frac{1}{18432},\, ...\right)
\end{align}
As the coefficients are stronly decreasing, in real calculations we can restrict ourselves to use only first two of them, i.e. we can make
the approximation of $I_0(x) = x$ and $I_1(x) = x^2/2$.

We can also calculate the elliptic flow, the second Fourier coefficient of the invariant single particle momentum distribution,
similarly to ref.~\cite{Csanad:2009wc}. Result is:
\begin{align}
&v_2(p_t) = \sum_{n=0}^{\infty}\frac{(2\pi)^{\frac{3}{2}}\sqrt{\rho_x\rho_y\rho_z}}{N_1(p_t)}\tau_0^4T_0
\left(\frac{p_t}{T_0}\right)^{\frac{3-4\kappa}{3}}\frac{\kappa}{3}\frac{B^n}{A^{n+\frac{4\kappa}{3}-\frac{3}{2}}}\nonumber\\
&\Bigg\lbrace  \left[ \frac{(\rho_x-1)^2+(\rho_y-1)^2}{8}(a_{0n}+2a_{1n}+a_{2n})- \right. \nonumber \\
& \left. - \frac{a_{0n}-2a_{1n}+a_{2n}}{4} \right]\left.\Gamma\left(n + \frac{4\kappa}{3}-\frac{3}{2}, A\frac{p_t}{T_0}\xi^{\frac{3}{\kappa}} \right) \right|_1^i + \nonumber \\
& \left. \frac{\rho_x^2 + \rho_y^2 + \rho_z^2}{2} \,a_{1n} A \left.\Gamma\left(n + \frac{4\kappa}{3}-\frac{5}{2}, A\frac{p_t}{T_0}\xi^{\frac{3}{\kappa}} \right) \right|_1^i \right\rbrace\label{e:v2}
\end{align}
Here the $I_2$ modified Bessel function had to be introduced as well, with coefficients of
$a_2 = \left(0,0,1/8, 0, 1/96, 0, 1/3072, 0,\, ...\right)$.

\section{Comparison to PHENIX data}
Hadronic data were described with this model in ref.~\cite{Csanad:2009wc}, determining freeze-out (final state) parameters
(expansion rates, freeze-out proper-time and freeze-out temperature). We use the parameters of the hadronic fit and leave only the remaining as free
parameters. The free parameters will be $\kappa$ (the equation of state parameter) and $t_i$, the initial time of the evolution.

\begin{figure}
 \begin{center}
 \includegraphics[width=0.81\textwidth]{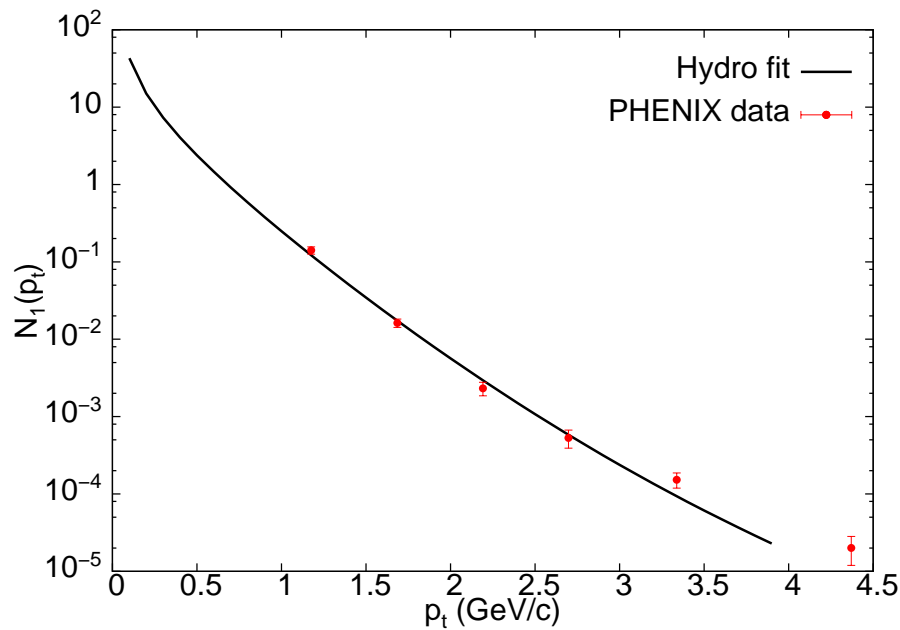}
 \end{center}
 \caption{Invariant transverse momentum of direct photons from our hydro model. The model validity goes until roughly 3 GeV in transverse momentum.}\label{f:N1pt}
\end{figure}

We use direct photon data from the PHENIX Collaboration~\cite{Adare:2008fqa}, measured in $\sqrt{s_{NN}}=$200 GeV Au+Au collisions.
Note that we also utilized a normalizing factor to describe the data. Fit parameters are summarized in table~\ref{t:param}.
The fit itself is shown on fig.~\ref{f:N1pt}. The equation of state result is $\kappa=7.7\pm0.7$, or alternatively, using $\kappa=1/c_s^2$:
\begin{align}
c_s = 0.36\pm0.02
\end{align}
which is in nice agreement with both lattice QCD calculations~\cite{Borsanyi:2010cj} and experimental
results from hadronic data~\cite{Adare:2006ti,Lacey:2006pn}. Note however that the spectrum is not very
sensitive to the initial time as in early times the thermal photon emission is not in the region of the experimental data. 
Our model does not contain acceleration, it is a Hubble-flow type of model, but the initial acceleration does not play a large role in the 
thermal photon spectrum because of this insensitivity on the initial time. We
determined an ``interval of acceptability'' for $t_i$. The maximum value for $t_i$ within 95\% probability is 0.7 fm/$c$. This can then be used
to determine a lower bound for the initial temperature. Thus the initial temperature of the fireball (in its center) is:
\begin{align}
T_i > 519\pm12\textnormal{MeV}
\end{align}
at 0.7 fm/$c$. The uncertainty comes from the uncertainty of $\kappa$. This is in accordance with other models' $300-600$ MeV result~\cite{Adare:2008fqa}.

\begin{table}
\begin{center}
    \begin{tabular}{ | c | c | c | c | }
    \hline
    Parameter &  & Value & Type\\ \hline \hline
    Central FO temperatre & $T_0$ & $204$ MeV & fixed \\ \hline
    FO proper-time & $\tau_0$ & $7.7$ fm/c & fixed \\ \hline
    Eccentricity & $\epsilon$ & $0.34$ & fixed \\ \hline
    Transverse expansion & $u_t^2/b$ & $-0.34$ & fixed \\ \hline
    Longitudinal expansion & $\dot{Z_0^2}/b$ & $-1.6$ & fixed \\ \hline
    Equation of State & $\kappa$ & $7.7 \pm 0.7$ & free \\ \hline
    Initial time & $t_i$ & $0 - 0.7$ fm$/c$ & free \\ \hline
    \end{tabular}   
\end{center}
\caption{The first five fit parametes were taken from the hadronic fits of ref.~\cite{Csanad:2009wc} (see more details about the parameters therein).
         The EoS parameter $\kappa$ was fitted, while for the initial time we determined an interval of acceptability (with 95\% confidence).
         The $\chi^2$ of the fit was 7.0 here, the number of points was 5, so with two fitted parameters our confidence level is 7.2\%}\label{t:param}
\end{table}

\bibliographystyle{prlstyl}
\bibliography{../../../../master}

\end{document}